\documentclass[aps,prl,twocolumn]{revtex4}
\usepackage{amsmath}
\usepackage{graphicx}
\usepackage{psfrag}
\usepackage{subfigure}
\newcommand*{\lifepo}[0]{{\ensuremath{\mathrm{LiFePO}_4}}}

\newcommand*{\gh}[0]{\ensuremath{g_{\mathrm{hom}}}}
\newcommand*{\Gm}[0]{\ensuremath{G_{\mathrm{mix}}}}
\newcommand*{\Ri}[0]{\ensuremath{R_{\mathrm{ins}}}}
\newcommand*{\vect}[1]{\ensuremath{\mathbf{#1}}} 
\newcommand*{\uvect}[1]{\ensuremath{\mathbf{\hat{#1}}}} 
\newcommand*{\tens}[1]{\ensuremath{\mathbf{#1}}} 
\newcommand*{\unit}[2]{\ensuremath{#1\mathrm{\,#2}}} 
\newcommand*{\nab}[0]{\ensuremath{\vect{\nabla}}}
\newcommand*{\pdop}[2][ ]{\partial^{#1}{#2}}
\newcommand*{\pdiff}[3][]{\frac{\pdop[#1]{#2}}{\pdop{#3^{#1}}}}
\newcommand{\abs}[1]{\left\lvert#1\right\rvert} 

\begin{document}
\author{Damian Burch}
\affiliation{Department of Mathematics, Massachusetts Institute of Technology, Cambridge, MA 02139-4307, USA}
\author{Martin Z. Bazant}
\affiliation{Department of Chemical Engineering and Department of Mathematics, Massachusetts Institute of Technology, Cambridge, MA 02139-4307, USA}
\email{bazant@mit.edu}
\title{Size-dependent spinodal and miscibility gaps for intercalation in nano-particles}


\begin{abstract}
Using a recently-proposed mathematical model for intercalation dynamics in phase-separating materials [Singh, Ceder, Bazant, Electrochimica Acta 53, 7599 (2008)], we show that the spinodal and miscibility gaps generally shrink as the host particle size decreases to the nano-scale. Our work is motivated by recent experiments on the high-rate Li-ion battery material \lifepo; this serves as the basis for our examples, but our analysis and conclusions apply to any intercalation material.  We describe two general mechanisms for the suppression of phase separation in nano-particles: (i) a classical bulk effect, predicted by the Cahn-Hilliard equation, in which the diffuse phase boundary becomes confined by the particle geometry; and (ii) a novel surface effect, predicted by chemical-potential-dependent reaction kinetics, in which insertion/extraction reactions stabilize composition gradients near surfaces in equilibrium with the local environment. Composition-dependent surface energy and (especially) elastic strain can contribute to these effects but are not required to predict decreased spinodal and miscibility gaps at the nano-scale.
\end{abstract}

\maketitle


{\bf Introduction. }
Intercalation phenomena occur in many chemical and biological systems, such as graphite intercalation compounds~\cite{dresselhaus1981}, DNA molecules~\cite{richards2007}, solid-oxide fuel cell electrolytes~\cite{ohare_book}, and Li-ion battery electrodes~\cite{huggins_book}. The intercalation of a chemical species in a host compound involves the nonlinear coupling of surface insertion/extraction reaction kinetics with bulk transport phenomena.  It can therefore occur by fundamentally different mechanisms in nano-particles and molecules compared to macroscopic materials due to the large surface-to-volume ratio.  Intercalation dynamics can also be further complicated by phase separation kinetics within the host material. This poses a challenge for theorists, since phase transformation models have mainly been developed for periodic or infinite systems in isolation~\cite{ballufi_book}, rather than nano-particles driven out of equilibrium by surface reactions.

In this paper, we ask the basic question, ``Is nano different?'', for intercalation phenomena in phase-separating materials. Our analysis is based on a general mathematical model for intercalation dynamics recently proposed by Singh, Ceder, and Bazant (SCB)~\cite{singh_ceder_bazant_2008}. The SCB model is based on the classical Cahn-Hilliard equation~\cite{cahn_1961} with a novel boundary condition for insertion/extraction kinetics based on local chemical potential differences, including concentration-gradient contributions. For strongly anisotropic nano-crystals, the SCB model predicts a new mode of intercalation dynamics via reaction-limited nonlinear waves that propagate along the active surface, filling the host crystal layer by layer. Here, we apply the model to the thermodynamics of nano-particle intercalation and analyze the size dependence of the miscibility gap (metastable uniform compositions) and the spinodal region (linearly unstable uniform compositions) of the phase diagram.

Our work is motivated by Li-ion battery technology, which increasingly involves phase-separating nanoparticles in reversible electrodes. The best known example is  \lifepo, a promising high-rate cathode material~\cite{padhi_nanjundaswamy_goodenough_1997} that exhibits strong bulk phase separation~\cite{padhi_nanjundaswamy_goodenough_1997,andersson_kalska_haggstrom_2000,chen_song_richardson_2006}.  Experiments have shown that using very fine nano-particles ($< \unit{100}{nm}$) can improve power density~\cite{yamada_chung_hinokuma_2001,huang_yin_nazar_2001} and (with surface modifications) achieve ``ultrafast'' discharging of a significant portion of the theoretical capacity~\cite{kang_ceder_2009}.  Experiments also provide compelling evidence~\cite{chen_song_richardson_2006,laffont_2006,delmas_maccario_croguennec_2008,ramana_2009} for the layer-by-layer intercalation waves (or ``domino cascade''~\cite{delmas_maccario_croguennec_2008}) predicted by the SCB theory~\cite{singh_ceder_bazant_2008,burch_singh_ceder_2008}, in contrast to traditional assumption of diffusion limitation in battery modeling~\cite{doyle_1993,srinivasan_2004}.

There is also experimental evidence that the equilibrium thermodynamics of \lifepo\ is different in nano-particles.
Recently, Meethong et al.\ have observed that, as the crystal size decreases, the miscibility gap between the lithium-rich and lithium-poor phases in the material shrinks significantly \cite{meethong_huang_carter_2007} (i.e.\ the tendency for phase separation is reduced).  A suggested explanation is that smaller particles experience relatively larger surface effects, which has been supported by calculations with an elaborate phase-field model \cite{tang_huang_meethong_2008}, although without accounting for surface reaction kinetics.  The size-dependency of the miscibility gap is fairly strong in \lifepo, and the importance of the surface energy has been demonstrated.  However, it has also been seen experimentally that carbon coating can reduce the surface effects and prevent the surface-induced reduction of the miscibility gap \cite{zaghib_mauger_gendron_julien_2008}.

We will show that the SCB model suffices to predict that the spinodal and miscibility gaps both decrease as the particle size decreases.  The analysis reveals two fundamental mechanisms:  (i)~nano-confinement of the inter-phase boundary, and (ii)~stabilization of the concentration gradients near the surface by insertion/extraction reactions.  These effects are independent of surface energy models, and indeed are valid for any phase-separating intercalation system.

{\bf Model. }
We employ the general SCB model for intercalation dynamics---based on the Cahn-Hilliard equation with chemical-potential-dependent surface reactions---without any simplifying assumptions~\cite{singh_ceder_bazant_2008}. In particular, we do not specialize to surface-reaction-limited or bulk-transport-limited regimes or perform any depth averaging for strongly anisotropic crystals~\cite{singh_ceder_bazant_2008,burch_singh_ceder_2008}.  Our field of interest is $c(\vect{x},t)$, the local concentration of the intercalated diffusing species (e.g., Li in \lifepo). Let $\rho$ be the density of intercalation sites per unit volume in the system (e.g., occupied by Li ions or vacancies), assumed to be constant and independent of position and local concentration.   We take $c$ to be normalized by $\rho$, so it is non-dimensional and only takes values between $0$ and $1$ (e.g., in the local compound Li$_c$FePO$_4$).

We assume that the free energy of mixing in our model system is well-approximated by the Cahn-Hilliard functional \cite{cahn_hilliard_1958,cahn_1961,ballufi_book}
\begin{equation}\label{eq:Gmix}
\Gm[c] = \int_V \left[ \gh(c) + \frac{1}{2}(\nab c)^T\tens{K}(\nab c) \right]\,\rho dV \ .
\end{equation}
The function $\gh(c)$ is the free energy per molecule of a homogeneous system of uniform concentration $c$, which is non-convex in systems exhibiting phase separation.  The gradient penalty tensor $\tens{K}$ is assumed to be a constant independent of $\vect{x}$ and $c$.  Then the diffusional chemical potential (in energy per molecule) is the variational derivative of $\Gm$,
\begin{equation}\label{eq:mu}
\mu(\vect{x},t) = \pdiff{\gh(c)}{c} - \nab\cdot(\tens{K}\nab{c}) \ .
\end{equation}
The mass flux (in molecules per unit area per unit time) is given by the linear constitutive relation \cite{degroot_mazur_1984}
\begin{equation}\label{eq:J}
\vect{J}(\vect{x},t) = -\rho c\tens{M}\nab\mu \ ,
\end{equation}
where $\tens{M}$ is a mobility tensor (denoted $\tens{B}$ by SCB~\cite{singh_ceder_bazant_2008}).  Finally, the dynamics are governed by the mass conservation equation
\begin{equation}\label{eq:ch}
\pdiff{(\rho c)}{t} + \nab\cdot\vect{J} = 0 \ .
\end{equation}

For illustration purposes, we employ the regular solution model for the homogeneous free energy \cite{prigogine_defay}:
\begin{equation}\label{eq:gh_rs}
\gh(c) = ac(1-c) + k_BT\left[c\log c + (1-c)\log(1-c)\right] \ .
\end{equation}
The two terms give the enthalpy and entropy of mixing, respectively.  When numerical values are needed, we will use $a/k_BT=5$, which is in rough agreement at room temperature with measurements on \lifepo\ \cite{dodd_yazami_fultz_2006}. Of course, other models are possible, but for the intercalation of a single species in a crystal with bounded compositions $0< c < 1$, the homogeneous chemical potential $\mu_\mathrm{hom}(c)=\gh^\prime(c)$  must diverge in the limits $c\to 0^+$ and $c\to 1^-$ due to entropic contributions from particles and vacancies.  (This constraint is violated, for example, by the quartic $\gh(c)$ from Landau's theory of phase transitions, suggested in a recent paper on \lifepo~\cite{ramana_2009} following SCB.)

Note that $\tens{K} / k_BT$ has units of length-squared.  Since it is assumed that $\tens{K}$ is positive-definite, we may denote its eigenvalues by $k_BT\lambda_i^2$ for real, positive lengths $\lambda_i$.  In particular, when $\tens{K}$ is diagonal, we define $\lambda_i\equiv\sqrt{K_{ii}/k_BT}$.  When the system is phase-separated into high-$c$ and low-$c$ regions, these $\lambda_i$ are the length scales for the interphasial widths in the different eigendirections~\cite{cahn_hilliard_1958,ballufi_book}.  In \lifepo, experimental evidence \cite{chen_song_richardson_2006} suggests that one of these widths is about \unit{4}{nm} (though the $\lambda_i$ in the other two directions might be large---comparable to the particle size---as phase-separation in these directions is not believed to occur).  These are therefore the natural length scales for measuring the size of phase-separating nano-crystals.

Our system of equations is closed by the following boundary conditions on the surface of the nano-particle:
\begin{align}
\label{eq:bc_var} \uvect{n}\cdot(\tens{K}\nab c) &= 0 \\
\label{eq:bc_flux} \uvect{n}\cdot\vect{J} &= -\rho_sR
\end{align}
where $\uvect{n}$ is an outward unit normal vector.  Equation~\ref{eq:bc_var} is the so-called variational boundary condition, which is natural for systems without surface energies or surface diffusion and follows from continuity of the chemical potential at the surface.  Equation~\ref{eq:bc_flux} is a general flux condition enforcing mass conservation, where  $\rho_s$ is the surface density of intercalation sites, and $R$ is the net local rate of intercalant influx (insertion) across the boundary. In the classical Cahn-Hilliard (CH) model, no mass flux across the boundary is allowed, and thus $R=0$. For intercalation systems~\cite{singh_ceder_bazant_2008,burch_singh_ceder_2008}, we allow for a non-zero reaction rate $R$ depending on the local values of $c$ and $\mu$ and refer to this general set of equations as the Cahn-Hilliard-\emph{with-reactions} (CHR) system.

\begin{figure*}[t]
     \centering
     \subfigure[Mixing energies of steady-state, phase-separated solutions.]
     {
         \psfrag{xlabel}[tc][tc]{average concentration ($\frac{1}{V}\int_V c\,dV$)}
         \psfrag{ylabel}[bc][bc]{$\Gm / \rho V k_B T$}
         \psfrag{legend1}[ml][ml]{$L = 10\lambda$}
         \psfrag{legend2}[ml][ml]{$L = 20\lambda$}
         \psfrag{legend3}[ml][ml]{$L = 100\lambda$}
         \includegraphics{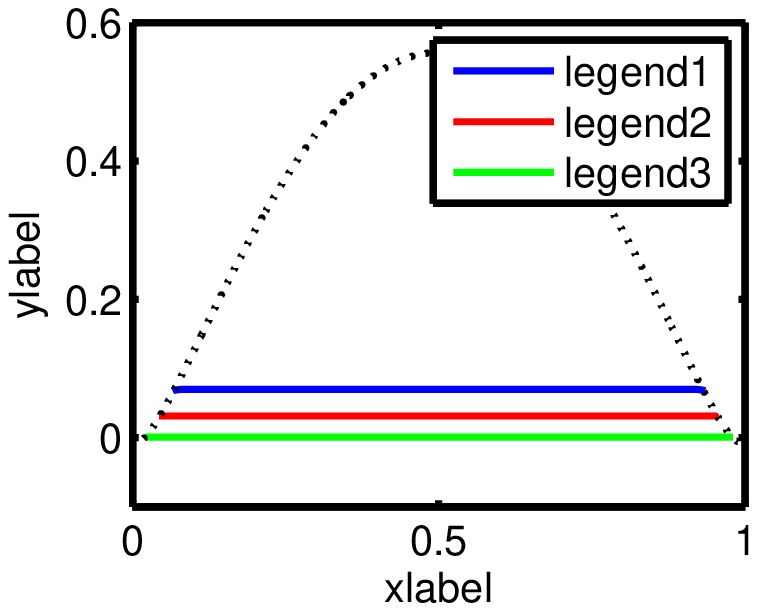}
         \label{fig:Gmix}
     }
     \subfigure[Width of the miscibility gap as a function of crystal size.]
     {
         \psfrag{xlabel}[tc][tc]{$L / \lambda$}
         \psfrag{ylabel}[bc][bc]{miscibility gap width}
         \includegraphics{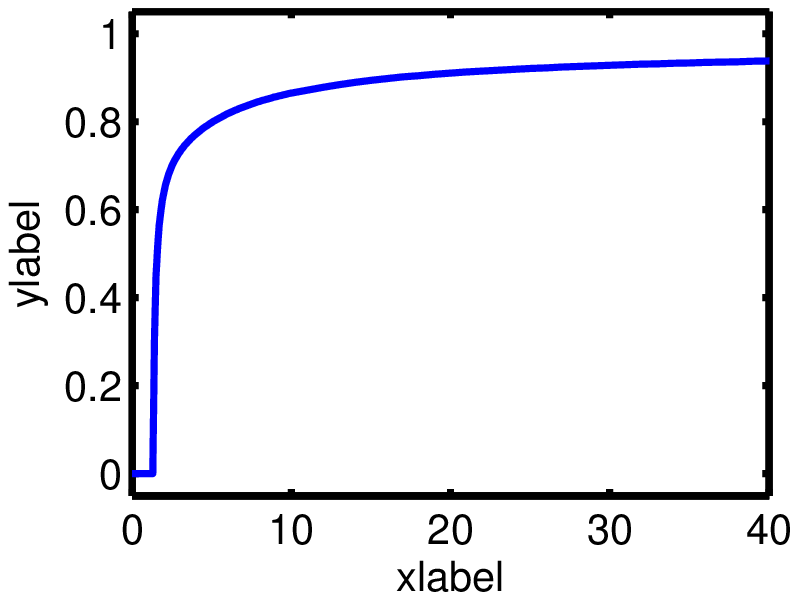}
         \label{fig:misc_gap}
     }
     \caption{The free energies in the first plot are given per intercalation site so that they are comparable across different crystal sizes.  The dotted line indicates the free energy of mixing per site for a uniform system of the given concentration.}
     \label{fig:miscibility}
\end{figure*}

For the current work (and indeed, for many of the conclusions reached by SCB~\cite{singh_ceder_bazant_2008}), the particular form of $R$ is unimportant. According to statistical transition-state theory in a concentrated solution, the net insertion rate is given by the difference of insertion and extraction rates, each having Arrhenius dependence on an (excess) chemical potential barrier. In order to satisfy de~Donder's equation \cite{prigogine_defay}, it must have the general form
\begin{equation}\label{eq:bc_rxn}
R = \Ri\left[1-\exp\left(\frac{\mu-\mu_e}{k_BT}\right)\right]
\end{equation}
where $\Ri$ is the rate for the insertion reaction. For thermodynamic consistency, this $\mu$ must be the same as the diffusional chemical potential used in the bulk equations, and $\mu_e$ is the external chemical potential of the intercalants in a reservoir phase outside of the particle (e.g., Li$^+$ in the electrolyte and $e^-$ in the metallic current collector of a Li-ion battery electrode); note that we are again assuming that the particle surface is energetically identical to the bulk.  If the reaction rates were controlled by electrostatic potential differences, for example, then $\Ri$ could include transfer coefficients and the interfacial voltage drop, and the familiar Butler-Volmer model for charge-transfer reactions \cite{bockris_reddy} would be recovered in the limit of a dilute solution. Following SCB~\cite{singh_ceder_bazant_2008}, we do not make any dilute solution approximation and keep the full CH expression for $\mu$ (\ref{eq:mu})---including the second derivative term---while assuming a uniform external environment at constant $\mu_e$. Although different models are possible for the chemical potential of the transition state~\cite{reaction}, we make the simple approximation of a constant insertion rate $\Ri$, consistent with particles impinging on the surface at constant frequency from the external reservoir. In that case, the composition dependence of $R$ enters only via the extraction rate.

{\bf The CH Miscibility Gap. }
Outside of the spinodal range, systems with uniform concentration fields are linearly stable.  However, if there exists a phase-separated solution with the same overall amount of our material but with a lower free energy, then the uniform system will only be metastable.  We will demonstrate that the miscibility range---the set of overall concentrations for which phase separation is energetically favorable---shrinks as the particle size decreases.

Unlike the spinodal, the miscibility gap cannot be studied analytically.  Instead, we must solve our original set of equations (\ref{eq:mu}--\ref{eq:ch}) numerically, looking for phase-separated systems with lower free energies than the uniform system with the same overall concentration.  We focus only on 1-dimensional systems, or equivalently 3-dimensional systems whose phase boundary is perpendicular to one of the eigendirections and whose concentration field is uniform in the other two directions.  Note that there is experimental \cite{chen_song_richardson_2006} and theoretical \cite{delmas_maccario_croguennec_2008} evidence that this is an accurate picture for the concentration field in \lifepo.  We will henceforth drop the subscripts on $\lambda$, and call $L$ the length of the system.

We begin by fixing a single crystal size.  For each value of the average concentration, we choose a corresponding initial condition, and we solve the Cahn-Hilliard equation (using a semi-discrete finite volume method and the no-flux boundary condition).  The system is stepped forward in time until the free energy reaches a minimum.  The resulting free energies of mixing for three different crystal sizes and a range of average concentrations are plotted in Fig.~\ref{fig:Gmix}.  Note that the curves do not extend across the entire $x$-axis.  This is because, for sufficiently extreme average concentrations, no initial conditions can be found which lead to a phase-separated steady state.  This suggests that such states do not exist, or that if they do exist they are not local minimizers of $\Gm$.  The phase-separated energy curves do extend slightly past the uniform curve, allowing us to estimate the endpoints of the miscibility gap.  The results suggest that the miscibility gap shrinks as the crystal size decreases.

In order to validate this hypothesis, we performed a more exhaustive search for phase-separated, steady-state solutions near the apparent miscibility endpoints.  This was done using the shooting method for boundary value problems to compute concentration fields satisfying \eqref{eq:mu} with $\mu=$ constant.  The resulting field that extremized the average concentration while still having a smaller free energy of mixing than the corresponding constant field was considered to be the boundary of the miscibility region.  The calculated miscibility gap widths over a range of crystal sizes are plotted in Fig.~\ref{fig:misc_gap}; they clearly support a shrinking miscibility gap.

There is a simple physical explanation for this behavior.  As discussed above, the interphase region will normally have a width on the order of $\lambda$.  The average concentration can only be close to $0$ or $1$ if this interphase region is close to a system boundary.  At this point, the average concentration can only become more extreme if the interphase region is compressed or disappears.  If it disappears, then we are left with a uniform system, and the average concentration is outside of the miscibility gap.  The other alternative, though, is expensive energetically due to the gradient penalty term in \eqref{eq:Gmix}.  Thus low-energy, phase-separated systems are limited geometrically to those concentrations in which the interphase region is (relatively) uncompressed between the crystal boundaries.  As the crystal size decreases, the limits imposed on the average concentration by the incompressibility of the interphase region becomes more and more severe, and thus the miscibility gap must shrink.

{\bf The CHR Spinodal Gap. }
The spinodal gap is the set of concentrations for which an initially-uniform system will spontaneously decompose through the exponential growth of infinitesimal fluctuations.  Thus, perturbation theory is the relevant mathematical tool, and we look for solutions to the CHR system of the form
\[
c(\vect{x},t) = c_0 + \epsilon c_1(\vect{x},t)
\]
where $c_0$ is a constant and $\epsilon$ is a small parameter.  If $c_0$ is truly a static solution to the CHR equations, then by \eqref{eq:bc_rxn}, $\mu_e$ must equal $\mu_0=\partial\gh(c_0)/\partial c$ at all points on the boundary of $V$.  The first-order system derived by linearizing about $c_0$ is then
\begin{subequations}\label{eq:perturbed}
\begin{align}
\mu_1(\vect{x},t)      &= \pdiff[2]{\gh(c_0)}{c}c_1 - \nab\cdot(\tens{K}\nab{c_1}) \\
\label{eq:pert_J} \vect{J}_1(\vect{x},t) &= -\rho c_0\tens{M}\nab\mu_1 \\
\pdiff{(\rho c_1)}{t}  &= -\nab\cdot\vect{J}_1
\end{align}
with the boundary conditions
\begin{align}
\uvect{n}\cdot(\tens{K}\nab c_1) &= 0 \\
\label{eq:pert_rxn} \uvect{n}\cdot\vect{J}_1 &= \frac{\rho_s\Ri}{k_BT}\mu_1 \ .
\end{align}
\end{subequations}
This is a fourth-order, linear system with constant coefficients.  Note that the exact same set of equations would result even had we taken $\tens{K}$ and $\tens{M}$ to be functions of $c$; the tensors above would only need to be replaced by the (still constant) values $\tens{K}(c_0)$ and $\tens{M}(c_0)$.

If we have an infinite system with no boundaries, then the Fourier ansatz $e^{i\vect{k}\cdot\vect{x}}e^{st}$ solves the above system if and only if it satisfies the dispersion relation
\begin{equation}\label{eq:dispersion}
s = -c_0(\vect{k}^T\tens{M}\vect{k})\left(\pdiff[2]{\gh(c_0)}{c}+\vect{k}^T\tens{K}\vect{k}\right) \ .
\end{equation}
Since $\tens{M}$ and $\tens{K}$ must be positive-semidefinite, $s$ will be non-positive whenever $\partial^2\gh(c_0) / \partial c^2 \ge 0$.  However, if $\partial^2\gh(c_0) / \partial c^2 < 0$, then the $c_0$ will be unstable to long-wavelength perturbations.  In particular, for the regular solution model \eqref{eq:gh_rs}, the criterion for linear stability becomes
\[
-2\frac{a}{k_BT} + \frac{1}{c_0(1-c_0)} \ge 0 \ .
\]
Thus a high enthalpy of mixing will promote instability of uniform systems with moderate concentrations.

\begin{figure}[t]
     \centering
     \psfrag{xlabel}[tc][tc]{$L / \lambda$}
     \psfrag{ylabel}[bc][bc]{spinodal gap width}
     \psfrag{legend1}[ml][ml]{$\mathcal{R} = \infty$}
     \psfrag{legend2}[ml][ml]{$\mathcal{R} = 0.1$}
     \psfrag{legend3}[ml][ml]{$\mathcal{R} = 0$}
     \includegraphics{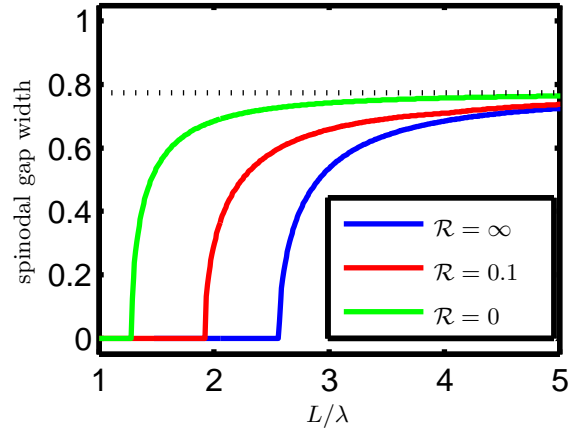}
     \caption{Width of the spinodal gap as a function of crystal size.  The dotted line indicates the width of the spinodal region for an infinite system.  The other three curves are given for different values of the non-dimensionalized reaction rate constant $\mathcal{R}\equiv (\rho_s/\rho\lambda)(\Ri/(D/\lambda^2))$, where $D=Mk_BT$ is the diffusion constant in the bulk.}
     \label{fig:spin_gap}
\end{figure}

If instead the system geometry is finite, then the boundary conditions will constrain the set of allowable wave vectors $\vect{k}$.  We again focus on one-dimensional systems for simplicity.  Then if the system occupies the line segment from $0$ to $L$, the general solution of the perturbed equations for the CH system ($\Ri=0$) is a sum of terms of the form
\[
c_1(x,t) = A\cos\left(\frac{n\pi}{L} x\right) e^{st}
\]
for any integer $n$.  The dispersion relation \eqref{eq:dispersion} still holds, but the wave number must equal $n\pi/L$ for integer values of $n$ in order to satisfy the boundary conditions.  In other words, we can no longer perturb the system with arbitrarily-long wavelengths.  The stability criterion is $\partial^2\gh(c_0) / \partial c^2 > -\pi^2 \lambda^2/L^2$.

For the regular solution model \eqref{eq:gh_rs}, the criterion for linear stability becomes
\[
-2\frac{a}{k_BT} + \frac{1}{c_0(1-c_0)} > -\pi^2 \lambda^2/L^2 \ .
\]
The spinodal region is defined as the range $(\alpha,1-\alpha)$ of unstable $c_0$ values.  It is easily verified that $\alpha$ is a decreasing function of $L$, i.e.\ that \emph{the spinodal range is more narrow for smaller crystals}.  Moreover, for sufficiently small values of $\lambda/L$, the above inequality is satisfied for all values of $c_0$, in which case there is no spinodal region at all.  These facts are demonstrated in \ref{fig:spin_gap}.  These results date back to Cahn's 1961 paper \cite{cahn_1961} and are known in the phase-field community.  However, it seems that their relevance for nano-particle composites---as in Li-ion batteries---has not yet been appreciated.

Moving beyond classical bulk models, we will now show that non-zero boundary reactions can further reduce the spinodal gap width.  Even the linear perturbed system of equations is no longer analytically tractable when $\Ri\ne 0$, and in particular, the wave numbers are no longer simply $n\pi/L$.  According to the dispersion relation \eqref{eq:dispersion}, every $s$ is associated with four wave numbers, and in general it takes a linear combination of all four such functions to satisfy the boundary conditions.  For any given $L$, $c_0$, and $s$, we may compute the four corresponding wave numbers $k_j$, and then look for a set of coefficients $A_j$ such that $\sum_{j=1}^4 A_je^{ik_jx}e^{st}$ solves the perturbed PDE and boundary conditions~\eqref{eq:perturbed}.  Because the system is linear and homogeneous, this can be reduced to finding a solution to some matrix equation $\tens{B}\vect{A}=\vect{0}$, which has solutions if and only if the determinant of the matrix $\tens{B}$ is $0$.

Therefore, for any given system size $L$ and reaction rate constant $\Ri$, we must numerically solve for the range of concentrations $c_0$ that admit solutions to the perturbed equations for at least one positive value of $s$.  Results of such computations are shown in \ref{fig:spin_gap}.  Notice that increasing the reaction rate constant reduces the spinodal gap.  Moreover, it was found numerically that increasing $\Ri$ tends to reduce the growth rate constant $s$.

These effects cannot be explained solely in terms of chemical potential perturbations near the boundary.  Instead, we must examine the nature of the allowable perturbations for different reaction rates.  For large values of $\Ri$, any non-zero $\mu_1$ at the boundaries causes large perturbations in the reaction fluxes by Eq.~\ref{eq:pert_rxn}.  In order for $J_1$ to be differentiable near the boundaries, we must also have large bulk fluxes nearby.  In general, this would require large concentration gradients, or equivalently short-wavelength perturbations.  But, as is clear from the dispersion relation~\eqref{eq:dispersion}, it is precisely the \emph{short}-wavelength perturbations which are rendered stable by the gradient penalty term (see Ref.~\cite{hilliard_1970} for an interesting discussion of this point).

More mathematically, suppose $\mu_1$ is non-zero at a boundary.  Then by \eqref{eq:pert_rxn}, $J_1$ must be non-zero there, which by \eqref{eq:pert_J} implies that $\mu_1$ must have a non-zero gradient.  Combining these two terms with our ansatz for $c_1$ yields the requirement
\[
\abs{c_0\left(\frac{\rho}{\rho_s}\right)\left(k_BTM\right)k} \sim \Ri \ .
\]
We therefore see that when $\mu_1$ is non-zero at a boundary, the wave number scales linearly with the reaction rate constant.  Again, large $\Ri$ would require large $k$, which are increasingly stable.

As the reaction rate increases, then, unstable perturbations satisfying \eqref{eq:perturbed} must have $\mu_1$ and $\nabla\mu_1$ close to $0$ near the boundaries.  However, this requires long-wavelength perturbations, and we have already shown that these will become increasingly stable as the crystal size shrinks.  Thus fast reaction rates will tend to stabilize small nano-particles.

Note, however, that for systems larger than about $2.5\lambda$, the spinodal gap does not disappear even for infinitely fast reactions.  This implies that there must exist infinitesimal perturbations to a uniform system which lead to phase separation without ever changing the diffusional chemical potential at the boundaries of the system.  This has been verified numerically by solving the full CHR system (\ref{eq:mu}--\ref{eq:bc_rxn}) in the limit $\Ri\rightarrow\infty$.  However, by limiting the spinodal decomposition to only occur via this small class of perturbations, higher reaction rates reduce the decomposition growth rate and the spinodal gap width.

Though we have used a specific mathematical model to derive these results, the conclusions are generally valid.  Regardless of the bulk model, a bounded system will only allow a discrete spectrum for its first-order perturbations.  The smallest admissible wave numbers will scale like $1/L$, and the system will suffer linear instability for more narrow ranges of concentrations as the system size shrinks.  Moreover, fast reaction rates at the boundaries require short wavelength perturbations, and such perturbations are energetically unfavorable when there is a diffuse interface between phases.

{\bf Other Effects. }
There are at least two important effects which we have excluded from our analysis.  First, we have intentionally ignored surface energies in order to demonstrate that purely bulk effects and reaction rates can cause shrinking spinodal and miscibility gaps.  However, surface energies could easily be accommodated.  For example, if the free energy of the system were to include a concentration-dependent surface tension between the particle and its environment,
\[
\Gm = G_\mathrm{mix,bulk} + \int_{\partial V} \gamma(c)\,dA \ ,
\]
then the variational boundary condition \eqref{eq:bc_var} would need to be replaced by
\[
\uvect{n}\cdot(\tens{K}\nab c) + \frac{1}{\rho}\frac{d\gamma}{dc} = 0 \ .
\]
This would change the analysis, but would not significantly affect the conclusions.

Perhaps a more serious omission for \lifepo\ in particular is elastic stress in the crystal due to lattice mismatches.  However, it has been demonstrated \cite{liam} that these effects can frequently be accommodated by simply decreasing the enthalpy-of-mixing parameter $a$.  Given our results above, elastic stress would therefore enhance the shrinking spinodal and miscibility gaps.

{\bf Conclusion. }
We have shown that intercalation phenomena in phase-separating materials can be strongly dependent on nano-particle size, even in the absence of contributions from surface energies and elastic strain.  In particular, the miscibility gap and spinodal gap both decrease (and eventually disappear) as the particle size is decreased to the scale of the diffuse interphase thickness.  Geometrical confinement enhances the relative cost of bulk composition gradients, and insertion/extraction reactions tend to stabilize the concentration gradients near the surface.  These conclusions have relevance for high-rate Li-ion battery materials such as \lifepo, but are in no way restricted to this class of materials.

{\bf Acknowledgements. }
This work was supported by the National Science Foundation under Contracts DMS-0842504 and DMS-0855011.

\bibliography{size_dep}

\end{document}